\magnification 1100
\baselineskip 13 pt


\def \v {\varphi}
\def \m {\mu}
\def \s {\sigma}
\def \b {\beta}
\def \an {a^{(n)}}

\def \rp {{\bf R}^+}
\def \vn {\varphi^{(n)}}
\def \l {\lambda}
\def \la {\lambda}
\def \a {\alpha}
\def \ul {\underline \lambda}

\def \sect#1{\bigskip  \noindent{\bf #1 } \medskip }
\def \subsect#1{\bigskip \noindent{\it #1 } \medskip}
\def \subsubsect #1{\bigskip \noindent{ #1 } \medskip}
\def \th#1#2{\medskip \noindent {\bf Theorem #1 }   \it #2 \rm \medskip}
\def \prop#1#2{\medskip \noindent {\bf Proposition #1 }   \it #2 \rm \medskip}
\def \cor#1#2{\medskip \noindent {\bf Corollary #1 }   \it #2 \rm \medskip}
\def \pf {\noindent  {\it Proof}.\quad }
\def \lem#1#2{\medskip \noindent {\bf Lemma #1 }   \it #2 \rm \medskip}
\def \ex#1{\medskip \noindent {\bf Example #1 }}

\def \noi {\noindent}
\def \hang {\hangindent 20 pt}

\def\sqr#1#2{{\vcenter{\vbox{\hrule height.#2pt\hbox{\vrule width.#2pt height#1pt \kern#1pt\vrule width.#2pt}\hrule height.#2pt}}}}

\def \square{\hfill\mathchoice\sqr56\sqr56\sqr{4.1}5\sqr{3.5}5}

\def \qed {$\square$ \medskip}

\centerline{\bf Valuation of Mortality Risk via the Instantaneous Sharpe Ratio:} \medskip
\centerline{\bf Applications to Life Annuities} \bigskip

\centerline {Version: 20 February 2008}
\bigskip

\noindent Erhan Bayraktar \hfill \break
\indent Department of Mathematics \hfill \break
\indent University of Michigan \hfill \break
\indent Ann Arbor, Michigan, 48109 \hfill \break
\indent erhan@umich.edu

\medskip

\noindent Moshe A. Milevsky \hfill \break
\indent Schulich School of Business \hfill \break
\indent York University \hfill \break
\indent Toronto, Ontario, M3J 1P3 \hfill \break
\indent milevsky@yorku.ca

\medskip

\noindent S. David Promislow \hfill \break
\indent Department of Mathematics and Statistics \hfill \break
\indent York University \hfill \break
\indent Toronto, Ontario, M3J 1P3 \hfill \break
\indent promis@yorku.ca

\medskip

\noindent Virginia R. Young \hfill \break
\indent Department of Mathematics \hfill \break
\indent University of Michigan \hfill \break
\indent Ann Arbor, Michigan, 48109 \hfill \break
\indent vryoung@umich.edu

\vfill \eject

\centerline{\bf Valuation of Mortality Risk via the Instantaneous Sharpe Ratio:} \medskip
\centerline{\bf Applications to Life Annuities} \bigskip

\noindent{\bf Abstract:}  We develop a theory for valuing non-diversifiable mortality risk in an incomplete market. We do this by assuming that the company issuing a mortality-contingent claim requires compensation for this risk in the form of a pre-specified instantaneous Sharpe ratio.  We apply our method to value life annuities.  One result of our paper is that the value of the life annuity is {\it identical} to the upper good deal bound of Cochrane and Sa\'{a}-Requejo (2000) and of Bj\"{o}rk and Slinko (2006) applied to our setting.  A second result of our paper is that the value per contract solves a {\it linear} partial differential equation as the number of contracts approaches infinity.  One can represent the limiting value as an expectation with respect to an equivalent martingale measure (as in Blanchet-Scalliet, El Karoui, and Martellini (2005)), and from this representation, one can interpret the instantaneous Sharpe ratio as an annuity market's price of mortality risk.

\medskip


\bigskip


\medskip


\sect{1. Introduction}

A basic textbook assumption on pricing life insurance and annuity contracts is that mortality risk is completely diversifiable and, therefore, should not be {\it priced} by capital markets in economic equilibrium. Indeed, under the traditional insurance pricing paradigm, the law of large numbers is invoked to argue that the standard deviation per policy vanishes in the limit and in practice a large enough portfolio is sufficient to eliminate mortality risk from the valuation equation. Although this only applies to a large number of claims that are independent random variables, it is commonly accepted that, in fact, typical real-world portfolios of life insurance and pension annuities satisfy the requisite assumptions.

Thus, for example, if actuaries estimate that 50\% of 65-year-olds in the year 2007 will live to see their 85th birthday in the year 2027, then a sufficiently homogeneous group of 65-year-olds should be charged $50$ cents per present value of \$1 dollar of a pure endowment contract (ignoring investment income). The underlying assumption is the existence of a well-defined population survival curve, which determines the fraction of the population living to any given time. This is also the probability of survival for any given individual within the group. Thus, if the insurance company sells enough of these policies and charges each policyholder the discounted value, on average the company will have enough to pay the \$1 to each of the survivors. Using the language of modern portfolio theory, the idiosyncractic risk, as measured by the standard deviation per policy, will go to zero if insurers sell enough policies, so the mortality risk is not priced.

However, a number of recent papers in the insurance and actuarial literature challenge this traditional approach within the framework of financial economics. This research falls under the general title of valuation of {\it stochastic mortality}.  In this paper, we provide a unique contribution to this literature by valuing stochastic mortality using techniques that have been traditionally applied to portfolio management.

Within the topic of stochastic mortality, recent papers by Milevsky and Promislow (2001),  Dahl (2004), DiLoernzo and Sibillo (2003), and Cairns, Blake, and Dowd (2006) proposed specific models for the evolution of the hazard rate.  Others, such as Blake and Burrows (2001), Biffis and Millossovich (2006), Boyle and Hardy (2004), Cox and Wang (2006), proposed and analyzed mortality-linked instruments. Few, if any, have focused on the actual equilibrium compensation for this risk.  Milevsky, Promislow, and Young (2006) provided some simple discrete-time examples, while Denuit and Dhaene (2007) used comonotonic methods to analyze this risk.  A practitioner-oriented paper by Smith, Moran, and Walczak (2003) used financial techniques to justify the valuing of mortality risk, although their approach is quite different from ours.

What this literature essentially argues is that uncertainty regarding the evolution of the instantaneous force of mortality induces a mortality dependence that cannot be completely diversified by selling more contracts.  This phenomenon induces a mortality risk premium that is valued by the market and whose magnitude will depend on a representative investor's risk aversion or demanded compensation for risk.  In other words, if there is a positive probability that science will find a cure for cancer during the next thirty years, aggregate mortality patterns will change.  Insurance companies must charge for this risk since it cannot be diversified.  The question we confront is: how much should they charge?

From a technical point of view, in this paper, we value mortality-contingent claims by assuming that the insurance company issuing the claims will be compensated for aggregate mortality risk via the instantaneous  Sharpe ratio of a suitably defined hedging portfolio.  Specifically, we assume that the insurance company chooses a hedging strategy that minimizes the local variance, as in Schweizer (2001a), then picks a target ratio $\a$ of net expected return to standard deviation, and finally determines the corresponding value for any given mortality-contingent claim that leads to this pre-determined $\alpha$.  See Bayraktar and Young (2008) for further discussion of this methodology from the standpoint of the mean-variance efficient frontier.

The main contribution of this paper is to develop a financial theory of valuing mortality-dependent contracts under stochastic hazard rates. We apply our method to value life annuities.  Arguably, the primary result of the paper is that the value per contract solves a {\it linear} partial differential equation as the number of contracts approaches infinity.  One can represent the limiting value as an expectation with respect to an equivalent martingale measure, and from this representation, one can interpret the instantaneous Sharpe ratio as the annuity market's price of mortality risk.

Although our valuation mechanism is very different--we use a continuous version of the standard deviation premium principle--it turns out that the value of the annuity corresponds to the upper good deal bound of Cochrane and Sa\'{a}-Requejo (2000) and of Bj\"{o}rk and Slinko (2006).  The no-arbitrage price interval of a contingent claim is a wide interval, and Cochrane and Sa\'{a}-Requejo (2000) find a reasonably small interval for prices by ruling out so-called good deals by putting a bound on the absolute value of the market price of risk of the Radon-Nikodym derivative of valuation measures with respect to the physical measure. The lower and upper prices for the resulting interval are obtained through  solving stochastic control problems, although this is not precisely stated in Cochrane and Sa\'{a}-Requejo (2000).  Our valuation method provides a new interpretation of the upper good deal bound of a life annuity as the value to a seller under the instantaneous Sharpe ratio $\a$ when the hedging strategy is chosen to minimize the local variance.  The lower good deal bound is the value to a buyer of a life annuity.

The remainder of this paper is organized as follows. In Section 2, we present our financial market, describe how to use the instantaneous Sharpe ratio to value the life annuity, and derive the resulting partial differential equation (PDE) that the value solves.  We also present the PDE for the value $\an$ of $n$ conditionally independent and identically distributed life annuity risks.

In Section 3, we compare our valuation method with two that are common in the literature.  Specifically, (1) we consider the good deal bounds of  Cochrane and Sa\'{a}-Requejo (2000) and of Bj\"{o}rk and Slinko (2006).  We show that for the issuer of a life annuity, our value is identical to the upper good deal bound; for the buyer of a life annuity, our value is identical to the lower good deal bound.  Therefore, we give an alternative derivation of the good deal bounds for our setting.  Then, (2) we consider indifference pricing via expected utility, as described in Zariphopoulou (2001), for example.   We observe that the relationship between our valuation method and indifference pricing is similar to the relationship between the standard deviation principle and variance principle in insurance pricing (Bowers et al., 1986).

In Section 4, we present several properties of $\an$ and find the limiting value of ${1 \over n} \an$.  We show that this limiting value solves a {\it linear} PDE and can be represented as an expectation with respect to an equivalent martingale measure.  Section 5 concludes the paper.

\sect{2. Instantaneous Sharpe Ratio}

In this section, we describe a life annuity and present the financial market in which the issuer of this contract invests.  We obtain the hedging strategy for the issuer of the life annuity.  We describe how to use the instantaneous Sharpe ratio to value the life annuity and derive the resulting PDE that the value solves.  We also present the PDE for the value $a^{(n)}$ of $n$ conditionally independent and identically distributed life annuity risks.

\subsect{2.1 $\,$ Mortality Model and Financial Market}

We use the stochastic model of mortality of Milevsky,  Promislow, and Young (2005).  We assume that the hazard rate $\la_t$ (or force of mortality) of an individual follows a diffusion process such that if the process begins at $\la_0 > \ul$ for some nonnegative constant $\ul$, then $\la_t > \ul$ for all $t \ge 0$.  From a modeling standpoint, $\ul$ could represent the lowest attainable hazard rate remaining after all causes of death such as accidents and homicide have been eliminated; see, for example, Gavrilov and Gavrilova (1991) and Olshansky, Carnes, and Cassel (1990).

Specifically, we assume that
$$
d\la_t = \mu(\la_t, t) \, (\la_t - \ul) \, dt + \sigma(t) \, (\la_t - \ul) \, dW^\la_t,  \quad \la_0 > \ul, \eqno(2.1)
$$
in which $\{ W^\la_t \}$ is a standard Brownian motion on a filtered probability space $(\Omega, {\cal F}, ({\cal F}_t)_{t \ge 0}, {\bf P})$.  The volatility $\s$ is either identically zero, or it is a continuous function of time $t$ bounded below by a positive constant $\kappa$ on $[0, T]$.  The drift $\m$ is such that there exists $\epsilon > 0$ such that if $0 < \la - \ul < \epsilon$, then $\mu(\la, t) > 0$ for all $t \in [0, T]$.  This condition on the drift ensures that $\l_t > \ul$ with probability 1 for all $t \in [0, T]$.  In general, we assume that $\m$ satisfies requirements such that (2.1) has a unique strong solution; see, for example, Karatzas and Shreve (1991, Section 5.2).  Note that if $\s \equiv 0$, then $\la_t$ is deterministic.

\ex{2.1} Suppose $\m(\l, t) = g + {1 \over 2} \s^2 + m \ln (\l_0 - \ul) + mg t - m \ln(\l - \ul)$ and $\s(t) \equiv \s$ in which $g \ge  0$, $m \ge 0$, and $\s > 0$ are constants.  The solution to (2.1) is given by
$$
\left\{ \eqalign{&\l_t = \ul + (\l_0 - \ul) \exp \left( g \, t +  Y_t \right), \cr
&dY_t = -m \, Y_t  \, dt + \s \, dW_t^\l, \quad Y_0 = 0,} \right. \eqno(2.2)
$$
a mean-reverting Brownian Makeham's law (Milevsky and Promislow, 2001).  Note that $Y_t = \s \int_0^t e^{-m(t-s)} \, dW_s^\l$, which reduces to $\s W_t^\l$ when $m = 0$.  As an aside, $\m(\l, t)$ satisfies the conditions in Assumption 4.1 below. \qed

Suppose an insurer issues a life annuity to an individual that pays at a continuous rate of \$1 per year until the individual dies or until time $T$, whichever occurs first.  Throughout this paper, we assume that the horizon $T$ is fixed.  In Section 2.2.2, to determine the value of the life annuity, we will create a portfolio composed of (1) the obligation to pay this life annuity, of (2) default-free, zero-coupon bonds that pay \$1 at time $T$, regardless of the state of the individual, and of (3) money invested in a money market account earning at the short rate.  Therefore, we require a model for bond prices, and we use a model based on the short rate and the bond market's price of risk.

The dynamics of the short rate $r$, which is the rate at which the money market increases, are given by
$$
dr_t = b(r_t, t) \, dt + c(r_t, t) \, dW_t, \eqno(2.3)
$$
in which $b$ and $c \ge 0$ are deterministic functions of the short rate and time, and $\{ W_t \}$ is a standard Brownian motion with respect to the probability space $(\Omega, {\cal F}, ({\cal F}_t)_{t \ge 0}, {\bf P})$, independent of $\{ W^\la_t \}$.  As for equation (2.1), we assume that $b$ and $c \ge 0$ are such that $r > 0$ almost surely and such that (2.3) has a unique solution; see, for example, Karatzas and Shreve (1991, Section 5.2).

From the principle of no-arbitrage in the bond market, there is a market price of risk process $\{ q_t \}$ for the bond that is adapted to the filtration generated by $\{W_t \}$; see, for example, Lamberton and Lapeyre (1996) or Bj\"ork (2004).  Moreover, the bond market's price of risk at time $t$ is a deterministic function of the short rate and of time; that is, $q_t = q(r_t, t)$.  Thus, the time-$t$ price $F$ of a default-free, zero-coupon bond that pays \$1 at time $T$ is given by
$$
F(r, t; T) = {\bf E^Q} \bigg[e^{-\int_t^T r_s ds} \, \bigg | \, r_t = r \bigg], \eqno(2.4)
$$
in which $\bf Q$ is the probability measure with Radon-Nikodym derivative with respect to $\bf P$ given by ${d{\bf Q} \over d{\bf P}} \big|_{{\cal F}_t} = \exp \left({-\int_0^t q(r_s, s) \, dW_s - {1 \over 2} \int_0^t q^2(r_s, s) \, ds} \right)$.  It follows that $\{W^Q_t \}$, with $W^Q_t = W_t + \int_0^t q(r_s, s) \, ds$, is a standard Brownian motion with respect to $\bf Q$.

From Bj\"ork (2004), we know that the price $F$ of the $T$-bond solves the following PDE:
$$
\left\{ \eqalign{& F_t + b^Q(r, t) F_r + {1 \over 2} c^2(r, t) F_{rr} - r F = 0, \cr
&F(r, T; T) = 1,} \right. \eqno(2.5)
$$
in which  $b^Q = b - q c$.  For the remainder of Section 2, we drop the notational dependence of $F$ on $T$ because $T$ is understood; whenever the bond pays at a time other than $T$, then we specify that time.

We can use the PDE (2.5) to obtain the dynamics of the $T$-bond's price $F(r_s, s)$, in which we think of $r_t = r$ as given and $s \in [t, T]$.  Indeed,
$$
\left\{ \eqalign{dF(r_s, s) &= (r_s \, F(r_s, s) + q(r_s, s) \, c(r_s, s) \, F_r(r_s, s)) ds + c(r_s, s) \, F_r(r_s, s) \, dW_s, \cr
F(r_t, t) &= F(r, t).} \right. \eqno(2.6)
$$
We use (2.6) in the next section when we develop the dynamics of a portfolio containing the obligation to pay the life annuity and a certain number of $T$-bonds.

\subsect{2.2 $\,$ Valuing via the Instantaneous Sharpe Ratio}

In this section, we first describe our method for valuing life annuities in this incomplete market.  Then, we fully develop the valuation for a single life annuity, that is, $n = 1$, and we present the valuation for the case when $n \ge 1$.

\subsubsect{2.2.1 $\,$ Recipe for valuation}

The market for insurance is incomplete; therefore, there is no unique pricing mechanism.  To value contracts in this market, one must assume something about how risk is ``priced.''  For example, one could use the principle of equivalent utility (see Zariphopoulou (2001) for a review) or consider the set of equivalent martingale measures (Blanchet-Scalliet, El Karoui, and Martellini, 2005; Dahl and M\o ller, 2006) to value the risk.  We use the instantaneous Sharpe ratio because of its analogy with the bond market's price of risk and because of the desirable properties of the resulting value; see Sections 3.1 and 4.  Also, in the limit as the number of contracts approaches infinity, the resulting value can be represented as an expectation with respect to an equivalent martingale measure.  Because of these properties, we anticipate that our valuation methodology will prove useful in pricing risks in other incomplete markets; for example, see Bayraktar and Young (2008).

Our method for valuing contingent claims in an incomplete market is as follows:

\smallskip

\item{A.}  First, define a portfolio composed of two parts:  (1) the obligation to underwrite the contingent claim (life annuity, in this case), and (2) a self-financing sub-portfolio of $T$-bonds and money market funds to (partially) hedge the contingent claim.

\smallskip

\item{B.}  Find the investment strategy in the bonds so that the local variance of the total portfolio is a minimum.  If the market were complete, then one could find an investment strategy so that the local variance is zero.  However, in an incomplete market, there will be residual risk as measured by the local variance.  This control of the local variance is called local risk minimization by Schweizer (2001a).

\smallskip

\item{C.}  Determine the value of the contingent claim so that the instantaneous Sharpe ratio of the total portfolio equals a pre-specified value.  This amounts to setting the drift of the portfolio equal to the short rate times the portfolio value {\it plus} the Sharpe ratio times the local standard deviation of the portfolio.  Cochrane and Sa\'a-Requejo (2000) and Bj\"ork and Slinko (2006) apply the idea of limiting the instantaneous Sharpe ratio to restrict the possible prices of claims in an incomplete market, and in Section 3.1, we further discuss how our method relates to theirs.  Also, Schweizer (2001b) derives a financial counterpart to the standard deviation premium principle.  His was a global standard deviation premium principle, whereas ours can be viewed as a local standard deviation premium principle.

\subsubsect{2.2.2 $\,$ Valuing life annuities}

Consider the time-$t$ value of the life annuity that pays at the continuous rate of \$1 per year until the individual dies or until time $T$, whichever comes first.  Denote the value of this annuity by $a = a(r, \la, t)$, in which we explicitly recognize that the value of the annuity depends on the short rate $r$ and the hazard rate $\la$ at time $t$.  (As an aside, by writing $a$ to represent the value of the annuity, we mean that the buyer of the annuity is still alive.  If the individual dies before time $T$, then the value of the policy jumps to 0.  Also, the value at time $T$ of the annuity is 0.)

As prescribed in Step A of Section 2.2.1, suppose the insurer creates a portfolio with value $\Pi_t$ at time $t$.  The portfolio contains (1) the obligation to underwrite the life annuity, with value $-a$, including the continuous flow of \$1 out of the portfolio to pay the life annuity, and (2) a self-financing sub-portfolio of $T$-bonds and money market funds with value $V_t$ at time $t$ to (partially) hedge the risk of the annuity.  Thus, while the individual is alive, $\Pi_t = -a(r_t, \la_t, t) + V_t$ with a continuous outflow of money at the rate of \$1 per year for $t \le T$.  Let $\pi_t$ denote the number of $T$-bonds in the self-financing sub-portfolio at time $t$, with the remainder, namely $V_t - \pi_t F(r_t, t)$, invested in a money market account earning the short rate $r_t$.

The annuity risk cannot be fully hedged because of the randomness inherent in the individual living or dying.  If the individual dies at time $t < T$, then the value of the annuity jumps to 0; therefore, the value of the portfolio $\Pi_t$ jumps from $-a(r_t, \la_t, t)  + V_t$ to $V_t$.  In other words, the value of the portfolio instantly changes by $a(r_t, \la_t, t)$.  Note that the portfolio that includes the value of the annuity is not self-financing, although the sub-portfolio with value $V_t$ is.

For $t \ge 0$, let $\tau(t)$ denote the future lifetime of the individual; that is, $t + \tau(t)$ is the time of death of the individual given that the individual is alive at time $t$.  Alternatively, $\tau(t) = \inf \{ u \ge 0: N(t + u) = 1  \big| N(t) = 0 \}$.  

\prop{2.1}{When $r_t = r$ and $\la_t = \la,$ the drift and local variance of the portfolio $\Pi_t$ are given by, respectively,
$$
\lim_{h \rightarrow 0} {1 \over h} \left({\bf E} \left(\Pi_{t+(h \wedge \tau(t))} - \Pi_t \big| {\cal F}_t \right) \right) = - {\cal D}^b a(r, \la, t) - 1 +  \pi_t \, q(r, t) \, c(r, t) \, F_r(r, t) + r \, \Pi,  \eqno(2.7)
$$
and
$$
\lim_{h \rightarrow 0} {1 \over h}  {\bf Var}(\Pi_{t+(h \wedge \tau(t))} | {\cal F}_t) =  c^2(r, t) (\pi_t \, F_r(r_s, s) - a_r(r, \la, t))^2 + \s^2(t) (\la - \ul)^2 a_\la^2(r, \la, t) +  \la a^2(r, \la, t).  \eqno(2.8)
$$}

\pf We first specify the dynamics of $a(r_t, \la_t, t)$ and $V_t$.  By It\^o's lemma (see, for example, Protter (1995)), the dynamics of the value of the life annuity are given as follows:
$$
\eqalign{&da(r_t, \la_t, t) = a_t \, dt + a_r \, dr_t + {1 \over 2} a_{rr} \, d[r, r]_t + a_\la \, d\la_t + {1 \over 2} a_{\la \la} \, d[\la, \la]_t \cr
& \quad = a_t \, dt + a_r (b \, dt + c \, dW_t) + {1 \over 2} a_{rr} \, c^2 \, dt + a_\la (\la_t - \ul) (\mu \, dt + \sigma \, dW^\la_t)  + {1 \over 2} a_{\la \la} \, \sigma^2 (\la_t - \ul)^2 dt,} \eqno(2.9)
$$
in which we suppress the dependence of the functions $a$, $b$, etc.\ on the variables $r$, $\la$, and $t$.  Also, $[r, r]$, for example, denotes the quadratic variation of $r$.  Because the sub-portfolio with value $V_t = \pi_t F(r_t, t) + (V_t - \pi_t F(r_t, t))$ is self-financing, its dynamics are given by
$$
\eqalign{dV_t &= \pi_t \, dF(r_t, t) + r_t (V_t - \pi_t F(r_t, t)) dt = (\pi_t \, q \, c \, F_r + r_t \, V_t) \, dt + \pi_t \, c \, F_r \, dW_t,} \eqno(2.10)
$$
in which the second equality follows from (2.6), and we again suppress the dependence of the functions on the underlying variables $r$ and $t$.

From the dynamics in equations (2.9) and (2.10), from the fact that the portfolio continually pays at the rate of \$1 while the individual is alive, and from the jump in the portfolio value when the individual dies (see the discussion preceding Proposition 2.1), it follows that the value of the portfolio at time $t + (h \wedge \tau(t))$ with $h > 0$, namely $\Pi_{t + (h \wedge \tau(t))}$, equals
$$
\eqalign{&\Pi_{t+(h \wedge \tau(t))} = \Pi_t - \int_t^{t+(h \wedge \tau(t))} da(r_s, \la_s, s) + \int_t^{t+(h \wedge \tau(t))} dV_s - \int_t^{t+(h \wedge \tau(t))} ds \cr
& \qquad \qquad \qquad + \int_t^{t+(h \wedge \tau(t))} a(r_s, \la_s, s) \, dN_s \cr
&= \Pi_t - \int_t^{t+(h \wedge \tau(t))} {(\cal D}^b a(r_s, \la_s, s) + 1) ds + \int_t^{t+(h \wedge \tau(t))} c(r_s, s) (\pi_s F_r(r_s, s) - a_r(r_s, \la_s, s)) \, dW_s \cr
& \quad - \int_t^{t+(h \wedge \tau(t))} \s(s) (\la_s - \ul) a_\la(r_s, \la_s, s) \, dW^\la_s + \int_t^{t+(h \wedge \tau(t))} a(r_s, \la_s, s) (dN_s - \la_s \, ds)  \cr
& \quad + \int_t^{t+(h \wedge \tau(t))}  \pi_s \, q(r_s, s) \, c(r_s, s) \, F_r(r_s, s) \, ds + \int_t^{t+(h \wedge \tau(t))} r_s \, \Pi_s \, ds,} \eqno(2.11)
$$
in which ${\cal D}^b$ is an operator defined on the set of appropriately differentiable functions on ${\bf R}^+ \times (\ul, \infty) \times [0, T]$ by
$$
{\cal D}^b v = v_t + b v_r + {1 \over 2} c^2 v_{rr} + \m (\la - \ul) v_\la + {1 \over 2} \s^2 (\la - \ul)^2 v_{\la \la} - (r + \la) v.  \eqno(2.12)
$$
$\{N_t\}$ denotes a counting process with stochastic parameter $\la_t$ at time $t$ that indicates when the individual dies.  Note that we adjust $dN_t$ in (2.11) so that $N_t - \int_0^t \la_s \, ds$ is a martingale.  The expressions in (2.7) and (2.8) follow from (2.11).  \qed

In this single-life case, the process $\Pi$ is ``killed'' when the individual dies.  If we were to consider the value $\an$ of $n$ conditionally independent and identically distributed lives (conditionally independent given the hazard rate), then $N$ would be a counting process with stochastic parameter $n \la_t$ until the first death such that $\Pi$ jumps by $\an - a^{(n - 1)}$ upon that death, and $N$ becomes a counting process with stochastic parameter $(n-1) \la_t$, etc.  We consider $\an$ later in this section and continue with the single-life case now.

As stated in Step B in Section 2.2.1, we choose the number of $T$-bonds $\pi_t$ to minimize the local variance of this portfolio, namely $\lim_{h \rightarrow 0} {1 \over h} {\bf Var}(\Pi_{t+(h \wedge \tau(t))} | {\cal F}_t)$, a dynamic measure of risk of the portfolio.  Note that this investment strategy is the same one advocated by Schweizer (2001a) in local risk minimization.  We have the following corollary of Proposition 2.1.

\cor{2.2}{The optimal investment strategy $\pi^*_t$ that minimizes the local variance is given by
$$
\pi^*_t = a_r(r_t, \la_t, t)/F_r(r_t, t),
$$
and under this assignment, the drift and local variance become, respectively,
$$
\lim_{h \rightarrow 0} {1 \over h} \left({\bf E} \left(\Pi_{t+(h \wedge \tau(t))} \big| {\cal F}_t \right) - \Pi \right) = - {\cal D}^{b^Q} a(r, \la, t) - 1 + r \Pi,  \eqno(2.13)
$$
and
$$
\lim_{h \rightarrow 0} {1 \over h} {\bf Var} \left(\Pi_{t+(h \wedge \tau(t))} \big| {\cal F}_t \right) = \s^2(t) (\la - \ul)^2 a_\la^2(r, \la, t) + \la a^2(r, \la, t), \eqno(2.14)
$$
in which the operator ${\cal D}^{b^Q}$ is defined in $(2.12)$ with $b$ replaced by $b^Q = b - qc$. $\square$}

Now, we come to valuing this annuity via the instantaneous Sharpe ratio, as in Step C of Section 2.2.1.  Because the minimum local variance  in (2.14) is positive, the insurer is unable to completely hedge the risk of the life annuity.  Therefore, the value should reimburse the insurer for its risk, say, by a constant multiple $\a \ge 0$ of the local standard deviation of the portfolio.  It is this $\a$ that is the instantaneous Sharpe ratio.  We restrict the choice of $\a$ to be bounded above by $\sqrt{\ul}$ to avoid values that violate the principle of no arbitrage; see Section 3.1 for further discussion of this point. 

To determine the value $a$ of the life annuity, we set the drift of the portfolio equal to the short rate times the portfolio {\it plus} $\a$ times the local standard deviation of the portfolio.  Thus, from (2.13) and (2.14), we have that $a$ solves the equation
$$
- {\cal D}^{b^Q} a - 1 + r \Pi = r \Pi + \a \sqrt{\s^2(t)(\la - \ul)^2 a_\la^2 + \la a^2}. \eqno(2.15)
$$
We summarize the above discussion in the following proposition.

\prop{2.3}{The value of the life annuity $a = a(r, \la, t)$ solves the non-linear PDE given by
$$
\left\{ \eqalign{&a_t + b^Q a_r + {1 \over 2} c^2 a_{rr} + \m (\la - \ul) a_\la + {1 \over 2} \s^2 (\la - \ul)^2 a_{\la \la} - (r + \la)a + 1 \cr
& \quad = - \a \sqrt{\s^2 (\la - \ul)^2 a_\la^2 + \la a^2}, \cr
& a(r, \la, T) = 0,} \right. \eqno(2.16)
$$
in which the terminal condition arises from the assumption that the annuity only pays until time $T$.  $\square$}

If we had been able to choose the investment strategy $\pi$ so that the local variance in (2.14) were identically zero (that is, if the risk were hedgeable), then the right-hand side of the PDE in (2.16) would be zero, and we would have a linear differential equation of the Black-Scholes type.

To end Section 2.2.2, we present the PDE solved by the value $\an$ of $n$ conditionally independent and identically distributed life annuity risks.  Specifically, we assume that all the individuals are of the same age and are subject to the same hazard rate as given in (2.1); however, given that hazard rate, the occurrences of death are independent.   As discussed in the paragraph following the proof of Proposition 2.1, when an individual dies, the portfolio value $\Pi$ jumps by $\an - a^{(n-1)}$.  By paralleling the derivation of (2.16), one obtains the following proposition.

\prop{2.4}{The value $\an = \an(r, \la, t)$ of $n$ conditionally independent and identically distributed life annuity risks solves the non-linear PDE given by
$$
\left\{ \eqalign{&\an_t + b^Q \an_r + {1 \over 2} c^2 \an_{rr} + \m (\la - \ul) \an_\la + {1 \over 2} \s^2 (\la - \ul)^2 \an_{\la \la} - r \an - n \la \left(\an - a^{(n-1)} \right) + n \cr
& \quad = - \a \sqrt{\s^2 (\la - \ul)^2 \left(\an_\la \right)^2 + n \la \left(\an - a^{(n-1)} \right)^2}, \cr
&  \an(r, \la, T) = 0} \right. \eqno(2.17)
$$
The initial value in this recursion is $a^{(0)} \equiv 0$, and the value $a$ defined by $(2.16)$ is $a^{(1)}$.}

Our model includes the special case for which $\a = 0$; this corresponds to the local risk minimization pricing and hedging model of Schweizer (2001a).  Therefore, our methodology is more general than that of local risk minimization.

\cor{2.5}{When $\a = 0$, then $\an = n \, a^{\a0}$ solves $(2.17),$ in which $a^{\a0}$ is given by
$$
a^{\a0} (r, \la, t) = \int_t^T F(r, t; s) \, {\bf E}^{\l, t}  \left[  e^{-\int_t^s \l_u du} \right] ds, \eqno(2.18)
$$
in which $F$ is the time-$t$ price of a default-free zero-coupon bond that pays $\$1$ at time $s$ as in $(2.4)$ with $T$ replaced by $s$. \qed}

\sect{3. Relationship with the Literature}

In Section 3.1, we show that our valuation of a life annuity for the seller is {\it identical} to the upper good deal bound of Cochrane and Sa\'a-Requejo (2000), generalized by Bj\"ork and Slinko (2006) to the case of a jump diffusion.  The lower good deal bound is obtained as the buyer's valuation of the life annuity.  Then, in Section 3.2, we compare our valuation with that given by indifference pricing via expected utility, as in Zariphopoulou (2001), for example.

\subsect{3.1 $\,$ Good Deal Bounds}

Our  work is this section is motivated by similar results of Bayraktar and Young (2008); see the remark at the end of their Section 2.5.  It is straightforward to show that one can write (2.16) as
$$
\left\{ \eqalign{&a_t + b^Q a_r + {1 \over 2} c^2 a_{rr} + {1 \over 2} \s^2 (\la - \ul)^2 a_{\la \la}  + 1 \cr
& \quad + \max_{\delta^2 + \la \gamma^2 \le \a^2, \, \gamma \ge -1} \big[ (\m + \delta \s) (\la - \ul) a_\la - (r + \la (1 + \gamma))a \big] = 0, \cr
& a(r, \la, T) = 0.} \right. \eqno(3.1)
$$
Note that the optimal values for $\delta$ and $\gamma$ are given by, respectively,
$$
\delta^* = {\alpha \sigma (\lambda - \underline \lambda) a_\lambda \over \sqrt{\sigma^2 (\lambda - \underline \lambda)^2 a^2_\lambda + \lambda a^2}},  \eqno(3.2)
$$
and
$$
\gamma^* = - {\alpha  a \over \sqrt{\sigma^2 (\lambda - \underline \lambda)^2 a^2_\lambda + \lambda a^2}}.  \eqno(3.3)
$$
Because we restrict $\a$ to lie between 0 and $\sqrt{\ul}$ inclusively, it follows that $\gamma^* > -1$ because $\l > \ul$.  Therefore, the restriction that $\gamma \ge -1$ in (3.1) is automatically satisfied at the maximum but we need it for the representation of $a$ in (3.4) below. 

From (3.1), it follows that $a$ can be represented as
$$
\eqalign{a(r, \la, t) &= \sup_{\{\{\delta_s, \gamma_s\}: \; \delta_s^2 + \la_s \gamma_s^2 \le \a, \; \gamma_s \ge -1, \; t \le s \le T \}} {\bf \bar E}^{r, \la, t} \left[ \int_t^T e^{-\int_t^s \left(r_u + \la_u \left(1 + \gamma_u \right) \right) \, du} \, ds \right] \cr
&= \sup_{\{\{\delta_s, \gamma_s\}: \; \delta_s^2 + \la_s \gamma_s^2 \le \a, \; \gamma_s \ge -1, \; t \le s \le T \}}  {\bf \bar E}^{r, \la, t} \left[ \int_t^T e^{-\int_t^s r_u \, du} \, {\bf 1}_{\{ \tau(t) > s \}}  \, ds \right].} \eqno(3.4)
$$
Here, $\{ r_t \}$ and $\{ \la_t \}$ follow the processes $d r_u = b^Q du + c \, d\bar W_u$ and $d \la_u = (\m + \delta \s)(\la_u - \ul) du + \s (\la_u - \ul) d \bar W^\la_u,$ respectively, with $\bar W_u = W_u + \int_0^u q(r_s, s) ds$ and $\bar W^\la_u = W^\la_u - \int_0^u \delta_u \, ds$.  The processes $\{ \bar W_t \}$ and $\{ \bar W^\la_t \}$ are independent standard Brownian motions with respect to the filtered probability space $(\Omega, {\cal F}, ({\cal F}_t)_{t \ge 0},  {\bf \bar P})$, in which ${d{\bf \bar P} \over d{\bf P}} \big|_{{\cal F}_t} = L_t$, with $L_t$ given by
$$
\left\{ \eqalign{&dL_t = L_{t-} \left( - q(r_t, t) \, dW_t + \delta_t \, dW_t^\la + \gamma_t(dN_t - \la_t \, dt) \right), \cr
&L_t  = 1.} \right.  \eqno(3.5)
$$
$\bf \bar E$ denotes expectation with respect to $\bf \bar P$.  The expression for $a$ in (3.4) is {\it identical} to the upper good deal bound given by Bj\"ork and Slinko (2006; Theorem 2.2).

As mentioned following the expression for $\gamma^*$ in (3.3), we have $\gamma^* > -1$  because $\a \le \sqrt{\ul}$.  This fact ensures that the expression in (3.4) is no greater than ${\bf \bar E}^{r, t} \big[\int_t^T F(r, t; s) \, ds \big]$, which is required by no arbitrage.  Also, note that $\gamma^*_s \ge -1$ is required in order for $\bf \bar P$ defined via (3.5) to be nonnegative.

Insurance actuaries are familiar with expressions similar to those in the expectation of (3.4) because in valuing life annuities, they effectively discount for interest and mortality often modifying the hazard rate to account for longevity risk (Wang, 1995).  To gain further understanding of  (3.4), we can write it as follows:
$$
a(r, \la, t) = \int_t^T F(r, t; s)  \sup_{\{\{\delta_s, \gamma_s\}: \; \delta_s^2 + \la_s \gamma_s^2 \le \a, \;  \gamma_s \ge -1, \; t \le s \le T \}} {\bf \bar E}^{\la, t} \left[ e^{-\int_t^s \la_u (1 + \gamma_u) du}  \right] \, ds, \eqno(3.6)
$$
in which we slightly abuse notation by writing ${\bf \bar E}$ here.

Because $\l_s(1 + \gamma_s) > 0$ with probability 1, we can think of it as a modified hazard rate in (3.6).  Note that if $\a = 0$, then $\gamma^* \equiv 0$ and $\delta^* \equiv 0$, and the expectation in the integral of (3.6) reduces to $\v^{\a0} (\la, t; s) := {\bf E}^{\l, t} \left[  \exp (-\int_t^s \l_u \, du) \right] $, the physical probability that a person survives from time $t$ to $s$, and the corresponding value of $a$ equals $a^{\a0}$ as in (2.18).

One can repeat the derivation in Section 2.2 to show that the value $a^b$ for the {\it buyer} of a life annuity solves (2.16) with $\a$ replaced by $-\a$.  Equivalently, $a^b$ solves (3.1) with {\it max} replaced by {\it min}.  Therefore, the buyer's value of the annuity is {\it identical} to the lower good deal bound of Bj\"ork and Slinko (2006; Theorem 2.2). It follows that we have the following bid-ask interval for life annuity values under our methodology, which we have just shown is equal to the one under good deal bounds:
$$
\left(\inf_{\{\delta_s, \gamma_s\}}{\bf \bar E^{r, \l, t}} \left[ \int_t^T e^{-\int_t^s r_u \, du} \, {\bf 1}_{\{ \tau(t) > s \}}  \, ds \bigg| {\cal F}_t \right], \; \sup_{\{\delta_s, \gamma_s\}}{\bf \bar E^{r, \l, t}} \ \left[ \int_t^T e^{-\int_t^s r_u \, du} \, {\bf 1}_{\{ \tau(t) > s \}}  \, ds \bigg| {\cal F}_t\right]\right), \eqno(3.7)
$$
in which we take the {\it sup}, respectively {\it inf}, over $\{\{\delta_s, \gamma_s\}: \; \delta_s^2 + \la_s \gamma_s^2 \le \a, \;  \gamma_s \ge -1, \; t \le s \le T \}$.  Therefore, we have provided an alternative derivation of the good deal bounds for our setting as the result of a local risk minimization investment strategy combined with a (local) risk loading expressed as a multiple of the local standard deviation of the portfolio.

The interval in (3.7) is a subinterval of the interval of no-arbitrage prices
$$
\left(\inf_{{\bf Q} \in {\cal M}}{\bf E^Q} \left[ \int_t^T e^{-\int_t^s r_u \, du} \, {\bf 1}_{\{ \tau(t) > s \}}  \, ds \Bigg| {\cal F}_t \right], \sup_{{\bf Q} \in {\cal M}}{\bf E^Q} \ \left[ \int_t^T e^{-\int_t^s r_u \, du} \, {\bf 1}_{\{ \tau(t) > s \}}  \, ds \Bigg| {\cal F}_t\right]\right), \eqno(3.8)
$$
in which ${\cal M}$ is the set of equivalent martingale measures.  Since this no-arbitrage pricing interval is too wide and practically useless, it was Cochrane and Sa\'{a}-Requejo (2000)'s idea to find a tighter subinterval.  From the representation of the seller's price in (3.4), one can see that our valuation method results in a coherent risk measure (Artzner et al., 1999).

\subsect{3.2 Indifference Pricing via Expected Utility}

Ludkovski and Young (2008) apply indifference pricing via expected utility to value pure endowments and life annuities in the presence stochastic hazard rates.  They use exponential utility so that the value of the insurance contracts are independent of the wealth of the seller.  With absolute risk aversion $\eta > 0$, one can follow the derivation of equation (2.41) in Ludkovski and Young (2008) to show that, in the setting of this paper, the indifference price $a^{IP} = a^{IP}(r, \la, t)$ solves the following PDE:
$$
\left\{ \eqalign{&a^{IP}_t + b^Q a^{IP}_r + {1 \over 2} c^2 a^{IP}_{rr} + \m (\la - \ul) a^{IP}_\la + {1 \over 2} \s^2 (\la - \ul)^2 a^{IP}_{\la \la} - (r + \la)a^{IP} + 1 \cr
& \quad = - {\eta \over 2 F} \s^2 (\la - \ul)^2 (a^{IP})_\la^2 - {\la F \over \eta} \left( e^{-\eta a^{IP}/F} - 1 + {\eta a^{IP} \over F} \right), \cr
& a^{IP}(r, \la, T) = 0.} \right. \eqno(3.9)
$$
Here $F = F(r, t; T)$.  Note that when $\eta = 0$, then $a^{IP} = a^{\a0}$, as defined in (2.18), which is also the price under the local risk minimization method of Schweizer (2001a).  The choice of $\a$ in (2.16) reflects the seller's risk aversion, just as the choice of $\eta$ in (3.9) reflects the seller's risk aversion in the setting of expected utility.

Parallel to equation (3.1), one can rewrite (3.9) as follows:
$$
\left\{ \eqalign{&a^{IP}_t + b^Q a^{IP}_r + {1 \over 2} c^2 a^{IP}_{rr} + {1 \over 2} \s^2 (\la - \ul)^2 a^{IP}_{\la \la} + 1 \cr
& \quad + \max_{\delta, \, \gamma} \bigg[ - {F \over 2 \eta} \delta^2 + {\la F \over \eta} (\gamma - (1 + \gamma) \ln(1 + \gamma))  \cr
& \qquad \qquad \quad + (\m + \delta \sigma) (\la - \ul) a^{IP}_\la -  (r + \la (1 + \gamma)) a^{IP}   \bigg] = 0, \cr
& a^{IP}(r, \la, T) = 0.} \right. \eqno(3.10)
$$
From (3.10), it follows that $a^{IP}$ can be represented as
$$
\eqalign{& a^{IP}(r, \la, t) \cr
& = \sup_{\{\delta_s, \gamma_s\}} {\bf \bar E}^{r, \la, t} \left[ \int_t^T e^{-\int_t^s \left(r_u + \la_u \left(1 + \gamma_u \right) \right) \, du} \, {F(r_s, s; T) \over \eta} \left( - {\delta_s^2 \over 2} + \la_s (\gamma_s - (1 + \gamma_s) \ln(1 + \gamma_s)) \right) \, ds \right].} \eqno(3.11)
$$
As in (3.4), $\{ r_t \}$ and $\{ \la_t \}$ follow the processes $d r_u = b^Q du + c \, d\bar W_u$ and $d \la_u = (\m + \delta \s)(\la_u - \ul) du + \s (\la_u - \ul) d \bar W^\la_u,$ respectively, with $\bar W_u = W_u + \int_0^u q(r_s, s) ds$ and $\bar W^\la_u = W^\la_u - \int_0^u \delta_u \, ds$.  Note that (3.11) is similar to the representation of $a$ in (3.4) with the constraint on the controls $\{ \delta_s, \gamma_s \}$ replaced by the penalty term $(F(r_s, s; T)/ \eta) \left( - \delta_s^2/2 + \la_s (\gamma_s - (1 + \gamma_s) \ln(1 + \gamma_s)) \right)$.

What follows is a formal discussion to show how the PDE in (3.9) is further related to the one in (2.16).  For $\eta a^{IP}/F$ small enough, one can ``approximate'' this PDE by expanding $e^{-\eta a^{IP}/F}$ and ignoring terms of the third power and higher.  When one does this, the resulting PDE is given by
$$
\left\{ \eqalign{&A_t + b^Q A_r + {1 \over 2} c^2 A_{rr} + \m (\la - \ul) A_\la + {1 \over 2} \s^2 (\la - \ul)^2 A_{\la \la} - (r + \la)A + 1 \cr
& \quad = - {\eta \over 2 F} \left( \s^2 (\la - \ul)^2 A_\la^2 + \l A^2 \right), \cr
& A(r, \la, T) = 0.} \right. \eqno(3.12)
$$
Compare this PDE with the one in (2.16).  Note that they are quite similar with $\a$ in (2.16) replaced by $\eta/(2F)$ in (3.12) and with the removal of the square root in (2.16).

Milevsky, Promislow, and Young (2005) show that valuing insurance risks via the instantaneous Sharpe ratio is closely related to the standard deviation principle, which is used by actuaries to price insurance (Bowers et al., 1986).  Similarly, Pratt (1964) shows that indifference pricing is closely related to the variance principle, under which the risk load is the square of the risk load under the standard deviation pricing principle.  Pratt's result is consistent with (3.12).  Thus, the valuation method that we propose in this paper is related to indifference pricing in a way parallel to how the standard deviation  principle is related to the variance principle, both of which are used in insurance pricing.

\sect{4. Properties of $\an$, the Value of $n$ Life Annuities}

To demonstrate important properties of $\an$, we rely on a comparison principle (Walter, 1970), and we present that in Appendix A.  In Section 4.1, we list some qualitative properties of $\an$.  Then, in Section 4.2, we prove the limiting result for ${1 \over n} \an$.

\medskip

\noindent{\bf Assumption 4.1} Henceforth, we assume that the volatility on the short rate $c$ satisfies the growth condition in the hypothesis of Theorem A.1 and that the drifts $b^Q$ and $\m$ satisfy the growth conditions in the hypothesis of Lemma A.2.  For later purposes (specifically for Property 3 in Proposition 4.1 below), we also assume that $\m_\la$ satisfies the growth condition $|\m_\la| (\l - \ul) + | \m| \le K \left( 1 + \left( \ln (\l - \ul) \right)^2  \right)$.  Also, we assume that $0 \le \alpha \le \sqrt{\ul}$, as stated in Section 3.1.

\subsect{4.1 $\,$ Qualitative Properties of $\an$}

In this section, we list and discuss properties of the risk-adjusted value $\an$ of $n$ life annuity contracts.  We state the following proposition without proof because its proof is similar to those in Milevsky, Promislow, and Young (2005).

\prop{4.1}{Under Assumption $4.1,$ $\an = \an(r, \l, t)$ satisfies the following properties on $G  = {\bf R}^+ \times (\ul, \infty) \times [0, T]$ for $n \ge 1:$  \smallskip
\item{$(1)$}  No arbitrage:  $0 \le \an \le n \int_t^T F(r, t; s) \, ds$.  \smallskip
\item{$(2)$}  Increasing in $n$:  $\an \ge a^{(n-1)}$.  \smallskip
\item{$(3)$}  Decreasing in $\la$: $\an_\la \le 0$.  \smallskip
\item{$(4)$}  Increasing in $\a$: Suppose $0 \le \a_1 \le  \a_2 \le \sqrt{\ul}$, and let $a^{(n), \a_i}$ be the solution of $(2.17)$ with $\a = \a_i,$ for $i = 1, 2$ and $n \ge 0$. Then, $a^{(n), \a_1} \le a^{(n), \a_2}$.  \smallskip
\item{$(5)$}  Lower bound:  $n a^{\a0} \le a^{(n), \a},$ in which $a^{\a0}$ is defined in $(2.18)$.  \smallskip
\item{$(6)$}  Decreasing in $\mu$:  Suppose $\m_1 \le \m_2$, and let $a^{(n), \m_i}$ denote the solution of $(2.17)$ with $\m = \m_i,$  for $i = 1, 2$.  Then, $a^{(n), \m_1} \ge a^{(n), \m_2}$.  \smallskip
\item{$(7)$}  Increasing in $\sigma$ if convex in $\la$:  Suppose $0 \le \s_1(t) \le \s_2(t)$ on $[0, T],$ and let $a^{(n), \s_i}$ denote the solution of $(2.17)$ with $\s = \s_i,$  for $i = 1, 2$.  If $a^{(n), \s_1}_{\la \la} \ge 0$ for all $n,$ or if $a^{(n), \s_2}_{\la \la} \ge 0$ for all $n,$ then $a^{(n), \s_1} \le a^{(n), \s_2}$.  \smallskip
\item{$(8)$}  Subadditive:  For $m, n$ nonnegative integers, $a^{(m+n)} \le a^{(m)} + a^{(n)}$.  \smallskip
\item{$(9)$}  Decreasing value per risk: ${1 \over n} \an$ decreases with respect to $n \ge 1$. \smallskip
\item{$(10)$}  Scaling: The value of $n$ annuities that pay at a rate of $k \ge 0$ is $k \, \an$.}

\noindent{\bf Remarks:}

\smallskip

\item{\bf 4.1}  Because the payoff under a life annuity is nonnegative, we expect its value to be nonnegative.  Also, if we hypothesize that the individuals will not die, then we get the upper bound given in Property 1.  In Section 4.2, we sharpen the bounds given in Property 1 considerably.

\smallskip

\item{\bf 4.2}  One uses Property 2 to prove Property 3.  However, Property 2 is interesting in its own right because it confirms our intuition that the marginal value of adding an additional policyholder, namely $\an - a^{(n-1)}$, is nonnegative.

\smallskip

\item{\bf 4.3}  Property 3 makes sense because we expect the value of life annuities to decrease as the hazard rate increases, that is, as individuals are more likely to die.

\smallskip

\item{\bf 4.4}  Property 4 states that as the parameter $\a$ increases, the risk-adjusted value $a^{(n), \a}$ increases.  This result justifies the use of the phrase {\it risk parameter} when referring to $\a$.  Also, by referring to the expression in (2.18), we see that Property 5 follows from Property 4.  Therefore, the lower bound of ${1 \over n} a^{(n), \a}$ (as $\a$ approaches zero) is the same as the lower bound of $a^\a = a^{(1), \a}$, namely, $a^{\a0}$.  We call the difference ${1 \over n} \an - a^{\a0} \ge 0$ the {\it risk charge} per annuity for a portfolio of $n$ life annuities.

\indent \indent \hang  Because ${1 \over n} \an - a^{\a0} \ge 0$, the strategy we propose results in a net profit on average, but this margin is necessary to compensate the issuer for the undiversifiable mortality risk, as in indifference pricing (Section 3.2).

\smallskip

\item{\bf 4.5}   In words, Property 6 tells us that as the {\it drift} of the hazard rate increases, then the value of the life annuities decreases.  This occurs for essentially the same reason that the value decreases with the hazard rate; see Property 3.  

\smallskip

\item{\bf 4.6}  Subadditivity is a reasonable property because if it did not hold, then buyers of annuities could  buy their annuities separately and thereby save money.  Subadditivity follows as a corollary from Property 9, which is interesting in itself, namely that the average value per risk decreases as the number of risks increases.  In other words, the risk charge per annuity, namely ${1 \over n} a^{(n)} - a^{\a0} \ge 0$, decreases as $n$ increases.  In the next section, we will determine the limiting value of the risk charge per annuity as $n$ goes to infinity, and from this result, we will subdivide the risk charge into the portion due to the finite size of the portfolio and the portion due to randomness of the hazard rate.

\smallskip

\item{\bf 4.7}  The same ten properties hold for the value of $n$ pure endowment risks (given in Appendix B), as shown in Milevsky, Promislow, and Young (2005), with Property 1 appropriately modified.

\medskip

These properties--which also hold in simpler, static settings (that is, for typical actuarial premium principles)--help to demonstrate that our valuation mechanism is reasonable.  What we gain in going to our more complicated setting is a dynamic pricing mechanism that compensates the issuer of the annuity for the undiversifiable mortality risk and shows the issuer how to hedge using a local risk minimization strategy.

\subsect{4.2 $\,$ Limiting Result for ${1 \over n} \an$}

In this section, we determine and interpret the limit $\lim_{n \rightarrow \infty} {1 \over n} \an$.  In the following two lemmas, we apply Theorem A.1 and Lemma A.2 to show that on $G  = {\bf R}^+ \times (\ul, \infty) \times [0, T]$, we have $n \int_t^T F(r, t; s) \, \b(\la, t; s) \, ds \le \an(r, \la, t) \le \int_t^T F(r, t; s) \, \vn(\la, t; s) \, ds$, in which $\b$ and $\vn$ are defined in (B.4) and (B.2), respectively, in Appendix B.  Thus, if we divide by $n$, we have that $\lim_{n \rightarrow \infty} {1 \over n} \an(r, \la, t) = \int_t^T F(r, t; s) \, \b(\la, t; s) \, ds$.  See Appendix C for the proofs of the two following lemmas.

\lem{4.2}{Under Assumption $4.1,$ we have
$$\an(r, \la, t) \le \int_t^T F(r, t; s) \, \vn(\la, t; s) \, ds,  \eqno(4.1)$$
on $G$ for $n \ge 0,$ in which $\vn$ is defined in $({\rm B}.2)$.}

From Lemma 4.2, we learn that if one were to treat $\vn$ as a modified expected number of survivors in a traditional actuarial computation of an annuity value, that is, the right-hand side of (4.1), then we overvalue the annuity.  Inequality (4.1) makes sense because $\an$ takes into account the reduction in overall risk when valuing an annuity (left-hand side of (4.1)) versus a series of essentially independent infinitesimal pure endowments (right-hand side of (4.1)). 

\lem{4.3} {Under Assumption $4.1,$ we have
$$n \int_t^T F(r, t; s) \, \b(\la, t; s) \, ds \le \an(r, \la, t), \eqno(4.2)$$
on $G$ for $n \ge 0,$ in which $\b$ is defined in $({\rm B}.4),$ or equivalently, in $({\rm B}.6)$.}

Lemma 4.3 implies that the value $\an$ of a life annuity is greater than $n$ times the value of a life annuity computed by using the limiting value of ${1 \over n} \vn$, namely $\b$.  In particular, $\an$ is greater than $n$ times the traditional actuarial value of a life annuity, namely $\int_t^T F(r, t; s) \, \v^{\a0}(\la, t; s) \, ds$ because $\b$ is greater than or equal to the physical probability of survival $\v^{\a0}$.  Recall from Appendix B that the value of $n$ conditionally independent and identically distributed pure endowments is given by $F \, \vn$.

We combine these two lemmas with a result from Milevsky, Promislow, and Young (2005)  to obtain the main result of this paper.

\th{4.4} {For $(r, \la, t) \in G,$
$$
\lim_{n \rightarrow \infty} {1 \over n} \an(r, \la, t) = \int_t^T F(r, t; s) \, \b(\la, t; s) \, ds. \eqno(4.3)
$$}

\pf From Milevsky, Promislow, and Young (2005), we know that the sequence $\left( {1 \over n} \vn \right)$ decreases uniformly to $\b$ on $(\ul, \infty) \times [0, s]$ for all $0 \le s \le T$. Also, recall from Property 9 in Proposition 4.1 that  ${1 \over n} \an$ decreases with respect to $n$.  Because $\int_t^T F \, \v \, ds \le \int_t^T \v \, ds \le {1 \over \ul - \a \sqrt{\ul}} < \infty$, it follows from the Lebesgue (dominated) convergence theorem (Royden, 1988, page 267) and Lemmas 4.2 and 4.3 that

$$\int_t^T F \, \b \, ds \le \lim_{n \rightarrow \infty} {1 \over n} \an \le \int_t^T F \lim_{n \rightarrow \infty} {1 \over n} \, \vn \, ds = \int_t^T F \, \b \, ds, \eqno(4.4)$$

\noindent which implies (4.3).  \qed

From the discussions accompanying $(2.4)$ and $({\rm B}.6),$ we can represent the limit in (4.3) as an expectation.

\cor{4.5}{For $(r, \la, t) \in G,$
$$
\lim_{n \rightarrow \infty} {1 \over n} \an(r, \la, t) = {\bf \hat E}^{r, \la, t} \left[ \int_t^T e^{-\int_t^s (r_u + \la_u) du} \, ds \right], \eqno(4.5)
$$
in which $\{ r_t \}$ and $\{ \la_t \}$ follow the processes $d r_u = b^Q du + c \, d\hat W_u$ and $d \la_u = (\m - \a \s)(\la_u - \ul) du + \s (\la_u - \ul) d \hat W^\la_u,$ respectively, with $\hat W_u = W_u + \int_0^u q(r_s, s) \, ds$ and $\hat W^\la_u = W^\la_u + \a u$.  The processes $\{ \hat W_t \}$ and $\{ \hat W^\la_t \}$ are independent standard Brownian motions with respect to the filtered probability space $(\Omega, {\cal F}, ({\cal F}_t)_{t \ge 0},  {\bf \hat P})$, in which ${d{\bf \hat P} \over d{\bf P}} \big|_{{\cal F}_t} = \exp \left( -\int_0^t q(r_s, s) \, dW_s - {1 \over 2} \int_0^t q^2(r_s, s) \, ds \right) e^{-\a W^\la_t - {1 \over 2} \a^2 t}$.  $\bf \hat E$ denotes expectation with respect to $\bf \hat P$.}

\noindent{\bf Remarks:}

\smallskip

\item{\bf 4.8}  The expression in (4.5) is identical to that of the value of a coupon bond that pays coupons continuously at the rate of \$1 under the discount rate of $r_t + \la_t$, instead of simply $r_t$ as for a bond that cannot default.  Therefore, one can think of $\la_t$ as the default rate.  For this reason, we anticipate that our methodology will prove useful in valuing credit risk derivatives.

\smallskip

\item{\bf 4.9}  The expression in (4.5) is an expectation with respect to a measure that is equivalent to the physical measure $\bf P$.  Therefore, in the limit, we obtain an arbitrage-free value for the life annuity.  Thereby, one can think of our valuation method as generalizing the one of Blanchet-Scalliet, El Karoui, and Martellini (2005).

\smallskip

\item{\bf 4.10}  If the hazard rate is deterministic, then in the limit, there is no mortality risk.  Specifically, if $\s \equiv 0,$ then $\lim_{n \rightarrow \infty} {1 \over n} \an = a^{\a0}$ on $G,$ in which $a^{\a0}$ is the net premium given in (2.18).  In other words, if the hazard rate is deterministic, then as the number of contracts approaches infinity, the value of a life annuity collapses to the discounted expected payout using the physical probability measure to value the mortality risk, regardless of the target value of the Sharpe ratio.  Alternatively, one can see that in the limit the average cash flow is certain, hence, the value becomes that of a ``certain'' (or non-life) annuity with a rate of discount modified to account for the rate of dying. 

\smallskip

\item{\bf 4.11} On the other hand, if the hazard rate is truly stochastic, then in the limit, mortality risk remains.  Specifically, if $\s$ is uniformly bounded below by $\kappa > 0$, then $\lim_{n \rightarrow \infty} {1 \over n} \an \ge a^{\a0}$ on $G,$ with equality only when $t = T$.

\smallskip

\item{\bf 4.12} We are now prepared to subdivide the risk charge per annuity into the portion due to the finite portfolio and the portion due to the stochastic hazard rate.  From the proof of Theorem 4.4, we know that ${1 \over n} \an(r, \la, t) \ge \int_t^T F(r, t; s) \, \b(\la, t; s) \, ds$, with equality in the limit.  Therefore, define ${1 \over n} \an - \int_t^T F \, \b \, ds$ as the risk charge (per risk) for holding a finite portfolio of $n$ annuities, and define $\int_t^T F \, \b \, ds - a^{\a0}$ as the risk charge for the stochastic hazard rate.  Thus, we have
$$
{1 \over n} \an - a^{\a0} = \left( {1 \over n} \an - \int_t^T F \, \b \, ds \right) + \left( \int_t^T F \, \b \, ds - a^{\a0} \right),  \eqno(4.6)
$$
in which the risk charge for the stochastic hazard rate is zero if $\sigma \equiv 0$ and is positive otherwise.

\smallskip

\item{\bf 4.13} The representation of the limiting value of ${1 \over n} \an$ in (4.3) is quite similar to a net single premium as defined in courses on Life Contingencies (Bowers et al., 1986).  Indeed, $F$ is the monetary discount function, and $\b$ plays the role of the survival probability.  Additionally, we can express the  limit $\int_t^T F(r, t; s) \, \b(\la, t; s) \, ds$ as the solution $p$ of the following PDE:

$$\left\{ \eqalign{&p_t + b^Q p_r + {1 \over 2} c^2 p_{rr} + (\m - \a \s) (\la - \ul) p_\la + {1 \over 2} \s^2 (\la - \ul)^2 p_{\la \la} - (r + \la)p + 1 = 0, \cr
& p(r, \la, T) = 0.} \right. \eqno(4.7)$$

\item{}  Note that $\a$ is analogous to the bond market's price of risk $q$ in (2.4); therefore, we refer to $\a$ as the (annuity) market's price of mortality risk.

\smallskip

\item{\bf 4.14} The proof of Theorem 4.4 suggests that we can use other reasonable models for bond pricing.  Additionally, one could use a different sort of default-free bond, such as a consol bond (or perpetuity).  In fact, in work not shown here, the authors obtained the same PDEs as in (2.16) and (2.17) when replacing the $T$-bonds with consol bonds.  The latter may be more appropriate in the setting of whole life annuities.  Alternatively, an insurer could continually roll money into longer-term bonds as the shorter-term bonds mature.

\smallskip

\item{\bf 4.15} By comparing Corollary 4.5 with the expression in (3.4) and the discussion afterwards, we see that as $n$ approaches infinity, the controls in the expectation in (3.4) are further restricted so that $\gamma_s \equiv 0$ with probability 1, that is, $\delta_s \equiv - \a$ with probability 1.

\smallskip

\item{\bf 4.16} Only in the limit does our valuation method result in a linear pricing rule, as evidenced by the linear PDE in (4.7) as compared with the non-linear PDE in (2.16).  Pricing via an equivalent martingale measure (EMM) results in a linear pricing rule; therefore, our general valuation method is not equivalent to choosing an EMM. 

\smallskip

\item{\bf 4.17} When valuing life insurance using the method of this paper, Young (2007) proves results parallel to those in Sections 4.1 and 4.2.  In particular, it is encouraging to learn that the limiting results are robust to the type of insurance contract (life insurance versus life annuities).  Also, in that paper, there is a numerical algorithm for valuing life insurance.  An interested reader can easily modify that numerical algorithm to value life annuities.

\sect{5. Conclusion}

A number of recent research papers focused attention on the valuation of longevity risk from a variety of  perspectives.  In this paper, we returned to basics and showed how the uncertainty in the hedging portfolio translates into a non-zero standard deviation per policy.  We draw upon the financial analogy of the Sharpe ratio to develop a methodology for valuing the non-diversifiable component of aggregate mortality risk.  Our main qualitative insight is that similar to the financial economic approach to analyzing stock market risk, the uncertainty embedded within mortality-contingent claims can be decomposed into idiosyncractic (diversifiable) and non-diversifiable components.

We developed a theoretical foundation for valuing mortality risk by assuming that the risk is ``priced'' via the instantaneous Sharpe ratio.  Because the market for life annuities is incomplete, one cannot assert that there is a unique price.  However, we believe that the price that our method produces is a valid one because of the many desirable properties that it satisfies and because it is the upper good deal bound of Cochrane and Sa\'a-Requejo (2000), generalized to the jump diffusion setting by Bj\"ork and Slinko (2006).  Therefore, we have the additional contribution of providing another derivation of the good deal bounds for the setting of this paper, one that includes a hedging strategy along with a actuarially-inspired risk loading in the form of the local standard deviation.

We also studied properties of the value of $n$ conditionally independent and identically distributed life annuities.  The value is subadditive with respect to $n$, and the risk charge per person decreases as $n$ increases.  We proved that if the hazard rate is deterministic, then the risk charge per person goes to zero as $n$ goes to infinity.  Moreover, we proved that if the hazard rate is stochastic, then the risk charge person is positive as $n$ goes to infinity, which reflects the fact that the mortality risk is not diversifiable in this case.  Additionally, in Remark 4.12, we decomposed the per-risk risk charge into the finite portfolio and stochastic mortality risk charges.


\sect{Appendix A. Comparison Principle}

In this appendix, we present a comparison principle from Walter (1970, Section 28) on which we rely extensively in the proofs of the properties  of $\an$ given in Section 4.  We begin by stating a relevant one-sided Lipschitz condition along with growth conditions.  Suppose a function $g = g(r, \la, t, v, p, q)$ satisfies the following one-sided Lipschitz condition:  For $v > w$,

$$g(r, \la, t, v, p, q) - g(r, \la, t, w, p', q') \le f_1(r, \la, t)(v - w) + f_2(r, \la, t) |p - p'| + f_3(r, \la, t) |q - q'|, \eqno({\rm A}.1)$$

\noindent with growth conditions on $f_1$, $f_2$, and $f_3$ given by

$$0 \le f_1(r, \la, t) \le K(1 + (\ln r)^2 + (\ln(\la - \ul))^2), \eqno({\rm A}.2a)$$

$$0 \le f_2(r, \la, t) \le K r (1 + |\ln r| + |\ln(\la - \ul)|), \eqno({\rm A}.2b)$$

\noindent and

$$0 \le f_3(r, \la, t) \le K (\la - \ul) (1 + |\ln r| + |\ln(\la - \ul)|), \eqno({\rm A}.2c)$$

\noindent for some constant $K \ge 0$, and for all $(r, \la, t) \in {\bf R}^+ \times (\ul, \infty) \times [0, T]$.  To prove properties of $\an$, we use the following comparison principle, which we obtain from Walter (1970, Section 28).  For the proof of a similar result, see Milevsky, Promislow, and Young (2005).

\th{A.1} {Let $G = {\bf R}^+ \times (\ul, \infty) \times [0, T],$ and denote by $\cal G$ the collection of functions on $G$ that are twice-differentiable in their first two variables and once-differentiable in their third. Define a differential operator $\cal L$ on $\cal G$ by
$${\cal L} v = v_t + {1 \over 2} c^2(r, t) v_{rr} + {1 \over 2} \s^2(t) (\la - \ul)^2 v_{\la \la} + g(r, \la, t, v, v_r, v_\la),  \eqno({\rm A}.3)$$
\noindent in which $g$ satisfies $({\rm A}.1)$ and $({\rm A}.2)$.  Suppose $v, w \in \cal G$ are such that there exists a constant $K \ge 0$ with $v \le e^{K \{(\ln r)^2 + (\ln(\la - \ul))^2\}}$ and $w \ge - e^{K \{(\ln r)^2 + (\ln(\la - \ul))^2\}}$ for large $(\ln r)^2 +(\ln(\la - \ul))^2$.  Also, suppose that there exists a constant $K \ge 0$ such that $c(r, t) \le K r$ for all $r > 0$ and $0 \le t \le T$. Then, if $($a$)$ ${\cal L} v \ge {\cal L} w$ on $G,$ and if $($b$)$ $v(r, \la, T) \le w(r, \la, T)$ for all $r > 0$ and $\la > \ul$, then $v \le w$ on $G$.}

\medskip

As a lemma for results to follow, we show that the differential operator associated with our problem satisfies the hypotheses of Theorem A.1.

\lem{A.2} {Define $g_n,$ for $n \ge 1,$ by
$$\eqalign{g_n(r, \la, t, v, p, q) &= b^Q(r, t) p + \m(\la, t) (\la - \ul) q - r v + n  - n \la \left(v - a^{(n-1)} \right) \cr
& \quad + \a \sqrt{\s^2(t) (\la - \ul)^2 q^2 + n \la \left(v - a^{(n-1)} \right)^2},} \eqno({\rm A}.4)$$
\noindent in which $a^{(n-1)}$ solves $(2.17)$ with $n$ replaced by $n-1$.  Then, $g_n$ satisfies the one-sided Lipschitz condition $({\rm A}.1)$ on $G$.  Furthermore, if $|b^Q(r, t) | \le K r (1 + |\ln r|)$ and $|\m(\la, t)| \le K (1 + |\ln(\la - \ul)|),$ then $({\rm A}.2)$ holds.}

\pf Suppose $v > w$,  then

$$\eqalign{&g_n(r, \la, t, v, p, q) - g_n(r, \la, t, w, p', q') \cr
& \quad = b^Q(r, t) (p - p') + \m(\la, t) (\la - \ul) (q - q') - (r + n \la)(v - w) \cr
& \qquad + \a \sqrt{\s^2(t) (\la - \ul)^2 q^2 + n \la \left(v - a^{(n-1)} \right)^2} \cr 
&\qquad - \a \sqrt{\s^2(t) (\la - \ul)^2 (q')^2 + n \la \left(w - a^{(n-1)} \right)^2} \cr
& \quad \le | b^Q(r, t) | |p - p'| + \left(|\mu(\la, t)| + \a \s(t) \right) (\la - \ul) |q - q'| - \left(r + n\la - \a \sqrt{n\la} \right) (v - w) \cr
& \quad \le | b^Q(r, t) | |p - p'| + \left(|\mu(\la, t)| + \a \s(t) \right) (\la - \ul) |q - q'|.} \eqno({\rm A}.5)$$

\noindent In the first inequality, we use the fact that if $A \ge B$, then $\sqrt{C^2 + A^2} - \sqrt{C^2 + B^2} \le A - B$.  For the second inequality, recall that $0 \le \a \le \sqrt{\ul}$.  Thus, (A.1) holds with $f_1(r, \la, t) = 0$, $f_2(r, \la, t) = | b^Q(r, t) |$, and $f_3(r, \la, t) =  |\mu(\la, t)| + \a \s(t)$.  Note that $f_2$ satisfies (A.2b) if $|b^Q(r, t) | \le K r (1 + |\ln r|)$, and $f_3$ satisfies (A.2c) if $|\m(\la, t)| \le K (1 + |\ln(\la - \ul)|)$.  $\square$

\sect{Appendix B. Limit Result for Valuing $n$ Pure Endowments}

In this appendix, we review some of the results of Milevsky, Promislow, and Young (2005).  Via the instantaneous Sharpe ratio, they value a pure endowment that pays 1 at time $s$ if an individual is alive at that time.  They also study properties of the time-$t$ value $P^{(n)} = P^{(n)}(r, \la, t; s)$ of $n$ conditionally independent and identically distributed pure endowment risks with $0 \le t \le s$.

By paralleling the derivation of (2.17), one can show that $P^{(n)}$ solves the non-linear PDE given by

$$\left\{ \eqalign{&P^{(n)}_t + b^Q P^{(n)}_r + {1 \over 2} c^2 P^{(n)}_{rr} + \m (\la - \ul) P^{(n)}_\la + {1 \over 2} \s^2 (\la - \ul)^2 P^{(n)}_{\la \la} - r P^{(n)} \cr & \quad - n \la \left(P^{(n)} - P^{(n-1)} \right) = - \a \sqrt{\s^2 (\la - \ul)^2 \left(P^{(n)}_\la \right)^2 + n \la \left(P^{(n)} - P^{(n-1)} \right)^2} \cr
& P^{(n)}(r, \la, s; s) = n,} \right. \eqno({\rm B}.1)$$

\noindent with $P^{(0)} \equiv 0$.

We can multiplicatively separate the variables $r$ and $\la$ in $P^{(n)}$.  Indeed, $P^{(n)}(r, \la, t; s) = F(r, t; s) \, \v^{(n)}(\la, t; s)$, in which $\v^{(n)}$ solves the recursion

$$\left\{ \eqalign{& \v^{(n)}_t + \m (\la - \ul) \v^{(n)}_\la +  {1 \over 2} \s^2 (\la - \ul)^2 \v^{(n)}_{\la \la} - n \la \left(\v^{(n)} - \v^{(n-1)} \right) \cr
& \qquad = - \a \sqrt{\s^2 (\la - \ul)^2 \left(\v^{(n)}_\la \right)^2 + n \la \left(\v^{(n)} - \v^{(n-1)} \right)^2}, \cr
& \v^{(n)}(\la, s; s) = n,} \right. \eqno({\rm B}.2)$$

\noindent with $\v^{(0)} \equiv 0$.  We can interpret $\v^{(n)}$ as a risk-adjusted expected number of survivors to time $s$ from the $n$ individuals alive at time $t$.  Indeed, Milevsky, Promislow, and Young (2005) show that $n {\bf E}^{\la, t} \left[ e^{-\int_t^s \la_u du} \right]  \le \v^{(n)} \le n e^{- \left(\ul - \a \sqrt{\ul} \right)(s - t)}$, and  $\v^{(n)}_\la \le 0$.  The main result of that paper is given in the following theorem:

\th{B.1} {Under Assumption $4.1,$ we have $$\lim_{n \rightarrow \infty}{1 \over n} P^{(n)}(r, \la, t; s) = F(r, t; s) \, \b(\la, t; s), \eqno({\rm B}.3)$$
uniformly on $\rp \times (\ul, \infty) \times [0, s]$. Here $\b$ is the solution of
$$\left\{ \eqalign{& \b_t + (\m - \a \s) (\la - \ul) \b_\la +  {1 \over 2} \s^2 (\la - \ul)^2 \b_{\la \la} - \la \b = 0, \cr
& \b(\la, s; s) = 1.} \right. \eqno({\rm B}.4)$$}

More precisely, Milevsky, Promislow, and Young (2005) show that

$$\b(\la, t; s) \le {1 \over n} \vn(\la, t; s), \eqno({\rm B}.5)$$

\noindent with equality in the limit.  Note that $\b$ solves a linear PDE; therefore, one can represent it as an expectation via the Feynman-Kac Theorem (Karatzas and Shreve, 1991):

$$\b(\la, t; s) = {\bf \tilde E}^{\la, t} \left[ e^{-\int_t^s \la_u du} \right], \eqno({\rm B}.6)$$

\noindent in which $\{ \la_t \}$ follows the process $d \la_u = (\m - \a \s)(\la_u - \ul) du + \s (\la_u - \ul) d \tilde W^\la_u$, with $\tilde W^\la_u = W^\la_u + \a u$.  The process $\{ \tilde W^\la_t \}$ is a standard Brownian motion with respect to the filtered probability space $(\Omega, {\cal F}, ({\cal F}_t)_{t \ge 0},  {\bf \tilde P})$, in which ${d{\bf \tilde P} \over d{\bf P}} \big|_{{\cal F}_t} = \exp \left({-\a W^\la_t - {1 \over 2} \a^2 t} \right)$.  $\bf \tilde E$ denotes expectation with respect to $\bf \tilde P$.

\sect{C. Proofs of Lemmas 4.2 and 4.3}

In this appendix, we supply the proofs of Lemmas 4.2 and 4.3.

\subsect{C.1 Proof of Lemma 4.2}

We proceed by induction.  The inequality is clearly true for $n = 0$ because both sides equal zero in that case.  Assume that inequality (4.1) holds for $n - 1$ in place of $n$ and show that it holds for $n$.  Define a differential operator $\cal L$ on $\cal G$ by (A.3) with $g = g_n$ in (A.4).  Because $\an$ solves (2.17), it follows that ${\cal L} \an = 0$, and

$$\eqalign{&{\cal L} \int_t^T F(r, t; s) \vn(\la, t; s) ds = -F(r, t; t) \vn(\la, t; t) + \int_t^T \left(F_t \vn + F \vn_t \right) ds \cr
& \quad + b^Q \int_t^T F_r \vn ds + {1 \over 2} c^2 \int_t^T F_{rr} \vn ds + \m (\la - \ul) \int_t^T F \vn_\la ds \cr
& \quad + {1 \over 2} \s^2 (\la - \ul) \int_t^T F \vn_{\la \la} ds - r \int_t^T F \vn ds - n \la \left( \int_t^T F \vn ds - a^{(n-1)} \right) + n \cr
& \quad + \a \sqrt{\s^2 (\la - \ul)^2 \left( \int_t^T F \vn_\la ds \right)^2 + n \la \left( \int_t^T F \vn ds - a^{(n-1)} \right)^2}.} \eqno({\rm C}.1)$$

\noindent Because $F(r, t; t) = 1$ and $\vn(\la, t; t) = n$, the first term on the right-hand side of (C.1) cancels the $n$ at the end of the third line.  Also, we use the fact that $F$ solves $F_t + b^Q(r, t) F_r + {1 \over 2} c^2(r, t) F_{rr} - r F = 0$ and that $\vn$ solves (B.2) to simplify (C.1) and thus obtain

$$\eqalign{&{\cal L} \int_t^T F(r, t; s) \vn(\la, t; s) ds = - n \la \left( \int_t^T F \v^{(n-1)} ds - a^{(n-1)} \right) \cr
& \qquad -  \a \int_t^T F \sqrt{\s^2 (\la - \ul)^2 \left( \vn_\la \right)^2 + n \la \left( \vn - \v^{(n-1)} \right)^2} ds \cr
& \qquad  + \a \sqrt{\s^2 (\la - \ul)^2 \left( \int_t^T F \vn_\la ds \right)^2 + n \la \left( \int_t^T F \vn ds - a^{(n-1)} \right)^2}.} \eqno ({\rm C}.2)$$

\noindent Now, ${\cal L} \left(\int_t^T F \vn ds \right) \le 0$ if

$$\eqalign{& \sqrt{\s^2 (\la - \ul)^2 \left( \int_t^T F \vn_\la ds \right)^2 + n \la \left( \int_t^T F \vn ds - a^{(n-1)} \right)^2} \cr
& \qquad  - \sqrt{n \la} \left( \int_t^T F \v^{(n-1)} ds - a^{(n-1)} \right) \cr
& \quad \le  \int_t^T F \sqrt{\s^2 (\la - \ul)^2 \left( \vn_\la \right)^2 + n \la \left( \vn - \v^{(n-1)} \right)^2} ds,} \eqno ({\rm C}.3)$$

\noindent where we use the assumption that $0 \le \a \le \sqrt{\ul}$, which in turn implies that $\a \sqrt{n \la} \le n \la$ for $n \ge 1$ and $\la > \ul$.

Recall that the Minkowski inequality tells us that if $0 < p < 1$, then $|| f ||_p + || g ||_p \le || f + g ||_p$, in which $|| f ||_p = \left( \int |f|^p \, d\nu \right)^{1/p}$ for some measure $\nu$, (Hewitt and Stromberg, 1965, page 192).  Let $p = 1/2$, $f = \s^2 (\la - \ul)^2 \left( \vn_\la \right)^2$, $g = n \la \left( \vn - \v^{(n-1)} \right)^2$, and $d\nu = F(r, t; s) \, ds$.  Then, the Minkowski inequality implies

$$\eqalign{&\left\{ \int_t^T F \sqrt{\s^2 (\la - \ul)^2 \left( \vn_\la \right)^2 + n \la \left( \vn - \v^{(n-1)} \right)^2} ds \right\}^2 \cr
& \quad \ge \s^2 (\la - \ul)^2 \left\{ \int_t^T F \vn_\la ds \right\}^2 + n \la \left\{ \int_t^T F \left( \vn - \v^{(n-1)} \right) ds \right\}^2.} \eqno ({\rm C}.4)$$

Thus, inequality (C.3) holds if

$$\eqalign{& \sqrt{\s^2 (\la - \ul)^2 \left( \int_t^T F \vn_\la ds \right)^2 + n \la \left( \int_t^T F \vn ds - a^{(n-1)} \right)^2} \cr
& \qquad - \sqrt{n \la} \left( \int_t^T F \v^{(n-1)} ds - a^{(n-1)} \right) \cr
& \quad \le \sqrt{\s^2 (\la - \ul)^2 \left( \int_t^T F \vn_\la ds \right)^2 + n \la \left( \int_t^T F \left( \vn - \v^{(n-1)} \right) ds \right)^2 }.} \eqno ({\rm C}.5)$$

\noindent If we define $B_\la = \s(t) (\la - \ul) \int_t^T F \, \vn_\la \, ds$, $A = \sqrt{n \la} \int_t^T F \, \vn \, ds$, $C = \sqrt{n \la} \int_t^T F \, \v^{(n-1)} \, ds$, and $D = \sqrt{n \la} \, a^{(n-1)}$, then (C.5) is equivalent to

$$\sqrt{B^2_\la + (A - D)^2} - (C - D) \le \sqrt{B^2_\la + (A - C)^2}.  \eqno ({\rm C}.6)$$

\noindent From Milevsky, Promislow, and Young (2005), we know that $A \ge C$.  From the induction assumption, we have that $C \ge D$.  Thus, the left-hand side of (C.6) is positive, and demonstrating  inequality (C.6) is a straightforward matter of squaring both sides and simplifying.

Thus, we have that ${\cal L} \left( \int_t^T F \, \vn \, ds \right) \le 0 = {\cal L} \an$.  Also, both $\int_t^T F \, \vn \, ds$ and $\an$ equal 0 when $t = T$.  It follows from Theorem A.1 and Lemma A.2 that $\an \le \int_t^T F \, \vn \, ds$ on $G$.  $\square$

\subsect{C.2 Proof of Lemma 4.3}

We proceed by induction.  The inequality is clearly true for $n = 0$ because both sides equal zero in that case.  Assume that inequality (4.2) holds for $n - 1$ in place of $n$ and show that it holds for $n$.  Define a differential operator $\cal L$ on $\cal G$ by (A.3) with $g = g_n$ in (A.4).  Because $\an$ solves (2.17), it follows that ${\cal L} \an = 0$, and

$$\eqalign{&{\cal L} \int_t^T n F(r, t; s) \, \b(\la, t; s) \, ds = -n F(r, t; t) \b(\la, t; t) + \int_t^T n \left(F_t \b + F \b_t \right) ds \cr
& \quad + b^Q \int_t^T n F_r \b ds + {1 \over 2} c^2 \int_t^T n F_{rr} \b ds + \m (\la - \ul) \int_t^T n F \b_\la ds \cr
& \quad + {1 \over 2} \s^2 (\la - \ul) \int_t^T n F \b_{\la \la} ds - r \int_t^T n F \b ds - n \la \left( \int_t^T n F \b ds - a^{(n-1)} \right) + n \cr
& \quad + \a \sqrt{\s^2 (\la - \ul)^2 \left( \int_t^T n F \b_\la ds \right)^2 + n \la \left( \int_t^T n F \b ds - a^{(n-1)} \right)^2}.} \eqno({\rm C}.7)$$

\noindent Because $F(r, t; t) = 1$ and $\b(\la, t; t) = 1$, the first term on the right-hand side of ({\rm C}.7) cancels the $n$ at the end of the third line.  Also, we use the fact that $F$ solves $F_t + b^Q(r, t) F_r + {1 \over 2} c^2(r, t) F_{rr} - r F = 0$ and that $\b$ solves (B.4) to simplify ({\rm C}.7) and thereby obtain

$$\eqalign{&{\cal L} \int_t^T n F(r, t; s) \b(\la, t; s) ds = n \la \left( a^{(n-1)} - (n - 1) \int_t^T F \b ds \right) \cr
& \qquad  + \a n \sqrt{\s^2 (\la - \ul)^2 \left( \int_t^T F \b_\la ds \right)^2 + {\la \over n} \left( \int_t^T n F \b ds - a^{(n-1)} \right)^2} \cr
& \qquad + \a n \s (\la - \ul) \int_t^T F \b_\la ds \ge 0 = {\cal L} \an,} \eqno({\rm C}.8)$$

\noindent in which the inequality follows from the induction assumption.  Also, both $\int_t^T F \, \b \, ds$ and $\an$ equal 0 when $t = T$.  It follows from Theorem A.1 and Lemma A.2 that $n \int_t^T F \, \b \, ds \le \an$ on $G$.  $\square$

\sect{Acknowledgments}

We thank two referees for carefully reading our manuscript and for providing very helpful suggestions.

\sect{References}

\noi \hang Artzner, P., F. Delbaen, J.-M. Eber, and D. Heath (1999), Coherent measures of risk, {\it Mathematical Finance}, 9 (3): 203-228.

\smallskip \noindent \hangindent 20 pt Bayraktar, E. and V. R. Young (2008), Pricing options in incomplete equity markets via the instantaneous Sharpe ratio, {\it Annals of Finance}, to appear, available at \hfill \break http://arxiv.org/abs/math/0701650

\smallskip \noindent \hangindent 20 pt Biffis, E. and P. Millossovich (2006), The fair value of guaranteed annuity options, {\it Scandinavian Actuarial Journal}, 2006 (1): 23-41.

\smallskip \noindent \hangindent 20 pt Bj\"ork, T. (2004), {\it Arbitrage Theory in Continuous Time}, second edition, Oxford University Press, Oxford.

\smallskip \noindent \hangindent 20 pt Bj\"ork, T. and I. Slinko (2006), Towards a general theory of good deal bounds, {\it Review of Finance}, 10: 221-260.

\smallskip \noindent \hangindent 20 pt Blanchet-Scalliet, C., N. El Karoui, and L. Martellini (2005), Dynamic asset pricing theory with uncertain time-horizon, {\it Journal of Economic Dynamics and Control}, 29: 1737-1764.

\smallskip \noindent \hangindent 20 pt Blake, D. and W. Burrows (2001), Survivor bonds: helping to hedge mortality risk, {\it Journal of Risk and Insurance}, 68: 339-348.

\smallskip \noi \hangindent 20 pt Bowers, N. L., H. U. Gerber, J. C. Hickman, D. Jones, and C. J. Nesbitt (1986), {\it Actuarial Mathematics}, Society of Actuaries, Schaumburg, Illinois.

\smallskip \noindent \hangindent 20 pt Boyle, P. P. and M. Hardy (2003), Guaranteed annuity options, {\it ASTIN Bulletin}, 33: 125-152.




\smallskip \noindent \hangindent 20 pt Cairns, A. J. G., D. Blake and K. Dowd (2006), A two-factor model for stochastic mortality with parameter uncertainty, {\it Journal of Risk and Insurance}, 73 (4): 687-718.


\smallskip \noindent \hangindent 20 pt Cochrane, J. and J. Sa\'a-Requejo (2000), Beyond arbitrage: good deal asset price bounds in incomplete markets, {\it Journal of Political Economy}, 108: 79-119.

\smallskip \noindent \hangindent 20 pt Cox, S. H., Y. Lin and S. Wang (2006), Multivariate exponential tilting and pricing implications for mortality securitization, {\it Journal of Risk and Insurance}, 73 (4): 719-736.

\smallskip \noindent \hangindent 20 pt Dahl, M. (2004), Stochastic mortality in life insurance: Market reserves and mortality-linked insurance contracts, {\it Insurance: Mathematics and Economics}, 35: 113-136.

\smallskip \noindent \hangindent 20 pt Denuit, M. and J. Dhaene (2007), Comonotonic bounds on the survival probabilities in the Lee-Carter model for mortality projection, {\it Journal of Computational and Applied Mathematics}, 203 (1): 169-176. 

\smallskip \noindent \hangindent 20 pt DiLorenzo, E. and M. Sibillo (2003), Longevity risk: measurement and application perspectives, working paper, Universita degli Studi di Napoli.



\smallskip \noindent \hangindent 20 pt Gavrilov, L. A. and N. S. Gavrilova (1991), {\it The Biology of Life Span: A Quantitative Approach}, Harwood Academic Publishers, New York.

\smallskip \noindent \hangindent 20 pt Hewitt, E. and K. Stromberg (1965), {\it Real and Abstract Analysis}, Springer-Verlag, Berlin.

\smallskip \noindent \hangindent 20 pt Karatzas, I. and S. E. Shreve (1991), {\it Brownian Motion and Stochastic Calculus}, second edition, Springer-Verlag, New York.

\smallskip \noindent \hangindent 20 pt Lamberton, D. and B. Lapeyre (1996), {\it Introduction to Stochastic Calculus Applied to Finance}, Chapman \& Hall/CRC, Boca Raton, Florida.


\smallskip \noindent \hangindent 20 pt Ludkovski, M. and V. R. Young (2008), Indifference pricing of pure endowments and life annuities under stochastic hazard and interest rates, {\it Insurance: Mathematics and Economics}, 42 (1): 14-30.

\smallskip \noindent \hangindent 20 pt Milevsky, M. A. and S. D. Promislow (2001), Mortality derivatives and the option to annuitize, {\it Insurance: Mathematics and Economics}, 29: 299-318.

\smallskip \noindent \hangindent 20 pt Milevsky, M. A., S. D. Promislow, and V. R. Young (2005), Financial valuation of mortality risk via the instantaneous Sharpe ratio: applications to pricing pure endowments, working paper, Department of Mathematics, University of Michigan, available at http://arxiv.org/abs/0705.1302

\smallskip \noindent \hangindent 20 pt Milevsky, M. A., S. D. Promislow, and V. R. Young (2006), Killing the law of large numbers: mortality risk premiums and the Sharpe ratio, {\it Journal of Risk and Insurance}, 73 (4): 673-686.



\smallskip \noindent \hangindent 20 pt Olshansky, O. J., B. A. Carnes, and C. Cassel (1990), In search of Methuselah: estimating the upper limits to human longevity, {\it Science}, 250: 634-640.

\smallskip \noindent \hangindent 20 pt Pratt, J. W. (1964), Risk  aversion in the small and in the large, {\it Econometrica}, 32: 122-136.

\smallskip \noindent \hangindent 20 pt Protter, P. (1995), {\it Stochastic Integration and Differential Equations}, Applications in Mathematics, 21, Springer-Verlag, Berlin.

\smallskip \noindent \hangindent 20 pt Royden, H. L. (1988), {\it Real Analysis}, third edition, Macmillan, New York.


\smallskip \noindent \hangindent 20 pt Schweizer, M. (2001a), A guided tour through quadratic hedging approaches, in {\it Option Pricing, Interest Rates and Risk Management}, E. Jouini, J. Cvitani\'c, and M. Musiela (editors), Cambridge University Press: 538-574.

\smallskip \noindent \hangindent 20 pt Schweizer, M. (2001b), From actuarial to financial valuation principles, {\it Insurance: Mathematics and Economics}, 28: 31-47.

\smallskip \noindent \hangindent 20 pt Smith, A., I. Moran, and D. Walczak (2003), Why can financial firms charge for diversifiable risk?, working paper, Deloitte Touche Tohmatsu, available at \hfill \break
http://www.casact.org/education/specsem/sp2003/papers/


\smallskip \noindent \hangindent 20 pt Walter, W. (1970), {\it Differential and Integral Inequalities}, Springer-Verlag, New York.

\smallskip \noindent \hangindent 20 pt Wang, S. S. (1995), Insurance pricing and increased limits ratemaking by proportional hazards transform, {\it Insurance: Mathematics and Economics}, 17 (1): 43-54.


\smallskip \noindent \hangindent 20 pt Young, V. R. (2007), Pricing life insurance under stochastic mortality via the instantaneous Sharpe ratio: Theorems and proofs, working paper, Department of Mathematics, University of Michigan, available at http://arxiv.org/abs/0705.1297


\smallskip \noindent \hangindent 20 pt Zariphopoulou, T., (2001), Stochastic control methods in asset pricing, {\it Handbook of \break Stochastic Analysis and Applications}, D. Kannan and V. Lakshmikantham (editors), Marcel Dekker, New York.

\bye